\documentclass[aps, amsmath,amssymb,twocolumn,superscriptaddress,showpacs,pre,reprint,nofootinbib]{revtex4-1}

\usepackage{amsmath}
\usepackage{natbib}
\usepackage{bm}
\usepackage{graphicx}  
\usepackage{graphics}
\usepackage{epsfig}
\usepackage{soul}
\usepackage{cancel}
\usepackage{hyperref}
\usepackage{bbold}
\usepackage{verbatim}
\usepackage{setspace}
\usepackage{color}
\definecolor{darkgreen}{rgb}{0,0.7,0.15}
\hypersetup{colorlinks,breaklinks,
            linkcolor=black,urlcolor=black,
            anchorcolor=black,citecolor=black}
\begin{document}

\newcommand{\EQ}[1]{Eq.~(\ref{eq:#1})}
\newcommand{\EQS}[2]{Eqs.~(\ref{eq:#1}) and (\ref{eq:#2})}
\newcommand{\FIG}[1]{Fig.~\ref{fig:#1}}
\newcommand{\TAB}[1]{Tab.~\ref{tab:#1}}
\newcommand{\REF}[1]{ref.~\citep{#1}}

\newcommand{\balst}[1]{\emph{#1}}
\newcommand{\xs}{\chi}
\newcommand{\ys}{\eta}
\newcommand{\ws}{\omega}
\newcommand{\rs}{\tilde{r}}
\newcommand{\dss}{\tilde{s}}
\newcommand{\x}{x}
\newcommand{\y}{y}
\newcommand{\w}{w} 
\newcommand{\rec}{r}
\newcommand{\ds}{s}
\newcommand{\mut}{U_b}
\newcommand{\ext}{\Psi}
\newcommand{\pfix}{P_{e}}
\newcommand{\psfix}{p_{e}}
\newcommand{\shift}{\theta}
\newcommand{\im}{{\rm Im}}
\newcommand{\csch}{{\rm csch}}
\newcommand{\tr}{{\rm Tr\,}}
\newcommand{\Wi}{{\rm Wi}\,}

\def\bal#1\eal{\begin{align}#1\end{align}}
\newcommand{\be}{\begin{equation}}
\newcommand{\ee}{\end{equation}}
\newcommand{\ba}[1]{\begin{array}{*{#1}{c}}}
\newcommand{\ea}{\end{array}}
\newcommand{\pr}{{\rm Prob}}
\newcommand{\pd}[2]{\frac{\partial #1}{\partial #2}}
\newcommand{\la}{\langle}
\newcommand{\ra}{\rangle}
\newcommand{\B}{\Theta}
\newcommand{\half}{\frac{1}{2}}
\newcommand{\OM}{\Omega}
\newcommand{\stp}{\sqrt{2\pi}}
\newcommand{\stps}{\sqrt{2\pi\sigma^2}}
\newcommand{\stovp}{\sqrt{2 \over \pi}}
\newcommand{\alert}[1]{\textcolor{red}{#1}}
\newcommand{\old}[1]{\textcolor{blue}{#1}}
\newcommand{\erf}{\,{\rm erf}}
\newcommand{\erfc}{\,{\rm erfc}}
\newcommand{\dd}{\,{\rm d}}
\newcommand{\curl}{\bm \nabla \times}
\newcommand{\divv}{\bm \nabla \cdot}
\newcommand{\grad}{\bm \nabla}
\newcommand{\const}{\,{\rm const}}

\title{Extreme value statistics of work done in stretching a polymer in a gradient flow}

\author{M.~Vucelja}
\email{vmarija@rockefeller.edu}
\affiliation{Center for Studies in Physics and Biology, The Rockefeller University, 1230 York Avenue, New York, NY 10065, USA}
\author{K.~S.~Turitsyn}
\affiliation{Department of Mechanical Engineering, Massachusetts Institute of Technology, Cambridge, MA, 02139, US}
\author{M.~Chertkov}
\affiliation{Theory Division \& Center for Nonlinear Studies at LANL and with New Mexico Consortium, Los Alamos,
NM 87545, US}
\date{\today}

\begin{abstract}
We analyze the statistics of work generated by a gradient flow to stretch a nonlinear polymer. We obtain the Large Deviation Function (LDF) of the work in the full range of appropriate parameters by combining analytical and numerical tools. The LDF shows two distinct asymptotes: "near tails" are linear in work and dominated by coiled polymer configurations, while "far tails" are quadratic in work and correspond to preferentially fully stretched polymers. We find the extreme value statistics of work for several singular elastic potentials, as well as the mean and the dispersion of work near the coil-stretch transition. The dispersion shows a maximum at the transition. 
\end{abstract}

\maketitle
\section{Introduction}
Most systems in nature are out of their equilibrium, dissipative and subject to external forces. Entropy production, heat production and work produced by an external force are common hallmarks of non-equilibrium systems characterizing the degree of the detailed balance violation. Recent intriguing results on production of entropy, work, as well as the statistics of the dissipation rate suggest new directions in non-equilibrium statistical physics. These results are stated in terms of various Fluctuation Theorems (FT), see e.g. \cite{ES,GC,K, LS,Maes,J2,Crooks} for theory and \cite{GCili,TCCP,SMS,VZC,VZCC,DJGPC,09GPCCG,Engel,Nickelsen2011aa} for applications to a variety of physical systems. A typical FT expresses the symmetry possessed by the probability distribution function (PDF) of the work accumulated over a long time. In this limit, the logarithm of the PDF is proportional to time, and the coefficient of proportionality is the Large Deviation Function (LDF).

A quantitative analysis of the LDF shape for linear systems has been reported in the literature, see e.g. \cite{TCCP,SMS,Engel}. In a nonlinear case the LDF is difficult to evaluate analytically. One obstacle is that the Gaussian Ansatz for the generating function of the work/entropy production (utilized in the linear stochastic problems) does not apply here. Farago gives the leading order estimate for the LDF for several pinning potentials \cite{2002Farago}, however does not discuss potentials due to singular forces (such as restitution forces of finitely extensible polymers). Also, straightforward numerical simulations are proven to be difficult in this regime, since even the vicinity of the global minimum of the LDF corresponds to rare events that are out of sampling reach for standard Monte-Carlo techniques. In this paper, we overcome these difficulties in deriving the extreme value statistics of the work done by stretching a polymer in a gradient flow. First we analyze the linear elasticity regime, similarly to \cite{TCCP}. Next we consider the other extreme - a regime where the polymers are preferentially stretched close to their maximal length by the external flow. The two cases give different asymptotics, connected by an intermediate region, which we obtain numerically, by implementing a rare-events sampling algorithm from \cite{GKP}. To the best of our knowledge, this approach of matching analytical estimates with numerics is novel. The method we use is general in that it is applicable for different nonlinear elasticities. We show that the LDF is sensitive to the type of the nonlinearity while in \cite{2002Farago} the LDF in leading order does not depend on the pinning potential. All of the potentials considered here have singularities, which makes it different from \cite{2002Farago}. We also obtain the mean and the dispersion of work. 

\section{A finitely extensible polymer in a gradient flow}
We study the statistics of work of a finitely extensible polymer subjected to a gradient flow and thermal fluctuations. The flow breaks the detailed balance and stretches the polymer. The work to stretch the molecule is stored as elastic energy, which later dissipates with fluctuations of the molecule's elongation. The whole system is in a non-equilibrium dynamical state, which is sustained by the energy flow from the fluid to the molecule and back. It is well documented in the literature \cite{87BAH} that even a minute amount of polymers is capable of generating significant non-Newtonian effects. Some of the most spectacular effects caused by anomalous stretching of polymers are rod climbing \cite{47Wei}, drag reduction \cite{48Tom} and elastic turbulence \cite{2000GS}. Analysis of the statistics of stretching of single polymers is a necessary prerequisite in order to grasp these phenomena.

We study the dumbbell polymer model in which, the polymer conformations are described solely by the end-to-end vector ${\bm r}(t)$. A more realistic polymer model would have a number of entropic springs connecting elements/beads and would also allow for hydrodynamic interactions between different beads. 
Numerical evidence suggest that the statistical nature of polymer chains is insensitive to the variation of bead number at sufficiently large and sufficiently weak stretching (more precisely Weissenberg number -- which we will define below) \cite{Watanabe:2010hv}. We consider the case where the polymer molecule is advected by an incompressible gradient flow, $\bm v = \sigma \bm r(t)$, correlated at length scales much larger the maximal polymer length $l$. The velocity gradient matrix, $\sigma = {\rm diag}(s, -s)$ is taken to be time-independent. The stochastic equation describing the balance of friction, elastic and thermal forces exerted on the polymer in the reference frame associated with its center of mass is
\begin{align}
\label{eq:langevin}
\zeta \left(\dot {\bm r}(t) - \bm v (\bm r (t))\right) = \bm F ({\bm r}(t)) + {\bm \xi}(t),
\end{align}
where $\bm F$ is the restitution force, ${\bm \xi}$ is the thermal noise and $\zeta$ is the friction coefficient \cite{87BAH}. We assume that the statistics of thermal forces is fully described by:  $\langle \xi_i(t)\rangle = 0$ and $\langle \xi_{i}(t)\xi_{j}(t')\rangle =  (2 \zeta/\beta) \delta _{ij}\delta(t-t')$. The potential energy can take different shapes, depending on the polymer stiffness, see e.g. \cite{MS, Warner}. Our main example is the Finitely Extendable Nonlinear Elastic (FENE) model with
\bal 
\bm F \equiv - \nabla U = -\gamma \frac{\bm r}{1-(r/ l)^2},
\eal but our analysis is general and we also apply it to the following elastic forces: $-\gamma \bm r/(1-(r/ l)^2)^{n}$ and $-\gamma \bm r/(1-(r/ l))^{n}$, where $n\in\mathbb{Z}^+$. The degree of polymer stretching can be expressed in terms of the Weissenberg number $\Wi \equiv s \tau$, which is defined as the product of the characteristic velocity gradient $s$ and the polymer relaxation time $\tau = \zeta /\gamma$. The value $\Wi=1$ separates the regime of  the ``coiled" phase of effectively linear elasticity, from the principally nonlinear phase, $\Wi >1$,  where the polymer is predominately stretched~\cite{73Lum}. The relaxation time to a steady state increases with the proximity of the coil-stretch transition~\cite{06CPV,08GS}, due to the abundance of different polymer configurations that contribute to the relaxation close to the transition. 

\section{The statistics of work done by the flow to stretch a polymer}

Work done by the flow to stretch the polymer fluctuates in time and it is given by
\bal
W[\bm r(\cdot)]\equiv \int_0^t\! \dd t' (\partial _{t'} + \bm v \cdot \nabla)U,
\eal 
where the material derivative takes into account the effects of the advection of the polymer by the external flow \cite{SMS, TCCP}. Langevin fluctuations translate into fluctuations of work, which are described by the PDF $\mathcal{P}(W|t)$.  At time $t$, which is parametrically larger than the correlation time $\tau_c\leq \{s^{-1},\tau\}$, one expects the PDF to take a large-deviation form: 
\bal
\mathcal{P}(w |t)\propto \exp[-\frac{t}{\tau_c} \mathcal{L}(w)],
\eal
where $w =\beta W\tau_c/t$ and $\mathcal{L}(w)$ is the LDF of the work produced over time $t$. Customarily in large deviation theory a rate function is defined as the tails of the cumulative distribution function of $w$ (see e.g. \cite{Ellis}), here $\mathcal{L}$ describes the tails of the PDF. The two rates at large enough times differ by logarithmic corrections $(\ln(t/\tau_c)$ terms). 

Our object of interest, $\mathcal{L}(w)$, is a convex function of its argument. To analyze it in detail we study the Laplace transform of $\mathcal{P}(W|t)$, also called the Generating Function (GF) of work, 
\bal
Z \equiv \left\langle e^{\eta\beta W[\bm r (\cdot)]}\right\rangle. 
\eal 
In the saddle point approximation we have that 
\bal
Z \simeq \exp\left[\frac{t}{\tau_c}(w_* \mathcal{L}'(w_*) - \mathcal{L}(w_*))\right] = e^{\frac{t}{\tau_c}\lambda(\eta) },
\eal 
where $\eta = \mathcal{L}'(w_*)$. The LDF and $\lambda(\eta)$ are the Legendre transforms of each other: $\mathcal{L}(w_*) = w_* \eta - \lambda(\eta)$. Below we will obtain $\lambda(\eta)$ and from there get the LDF. The Gallavotti-Cohen fluctuation theorem \cite{LS,GC,GC1995aa} implies the following relation $\mathcal{L}(w) = \mathcal{L}(-w) - w$, which is equivalent to $\lambda (\eta) = \lambda(-1-\eta)$. Hence in order to get $\lambda(\eta)$ for $\eta \in \mathbb{R}$, it is enough to look at $\eta > -\frac{1}{2}$. 

The ``standard" fluctuation theorem relates the probabilities of positive and negative entropy production in the same system. Here it is valid only if the flow and its time-inverse image are physically equivalent, i.e. coincide after properly chosen spatial rotation and inversion. Although all planar flows satisfy this, the condition is broken in a generic three-dimensional gradient flow. For example the "standard" FT is violated for a three-dimensional axially-symmetric elongational flow. Such a flow can be specified with velocity gradient matrix of the following form: $\mathrm{diag}(2s,-s,-s)$. Namely while such a flow with $s>0$ would deform a spherical blob of passive scalar (e.g. dye) into a one-dimensional filament, it's time-reversed copy ($s \to -s$) would turn the same blob into a two-dimensional "pancake". 
 
The GF $Z$ is conditioned on the initial $\bm r(0)$ and final point $\bm r(t)$. It can be formally expressed in terms of the path-integral in the polymer configuration space as
\bal
\label{eq:gen-function-path-int}
&Z = \int^{\bm r(t)} _{\bm r(0)}\! [{\cal D} {\bm r}(\cdot)] \exp\! \left[\! -S[\bm r(\cdot)]\!- \eta\beta\int ^t _0 \dd t' \bm v \cdot \bm F \right]\!,\!\!\!
\\
\label{eq:action}
&S \equiv - \frac{\zeta\beta}{4} \!\int ^t _0 \!\!\!\!\dd t'\! \!\left(\!\!\left(\dot{\bm r} - \bm v - \frac{\bm F}{\zeta}\!\right)^{\!2} \!\!+\!\frac{2}{\beta\zeta} \nabla \!\cdot\!\left( \!\bm v + \frac{\bm F}{\zeta}\!\right)\!\!\right)\!\!,\!\!\!\!
\eal 
where $S$ is the effective action \cite{92Kam,Engel}. From \EQ{gen-function-path-int} one obtains the Fokker-Planck equation (see e.g. \cite{CCJ})
\bal
\label{eq:FPZ}
\partial _t Z = - \nabla \cdot \left(\left(\frac{\bm F}{\zeta} +\bm v\right)\! Z\!\right) + \frac{\nabla ^2 Z}{\beta \zeta} - \eta \beta\zeta \bm v\cdot \frac{\bm F}{\zeta} Z.
\eal
It is convenient to make the variables dimensionless. Onwards the unit of temperature is  $(\gamma l^2/2)$, the unit of the polymer length is  $l$ and time is measured in units of $\tau$. 

We apply the substitution 
\bal
Y = \exp\left[-\frac{\Wi}{{T}}\int \dd \bm r \cdot \left(\bm v + \frac{\bm F}{\Wi}\right)\right]Z
\eal
to  \EQ{FPZ} and get a Schr\" odinger like equation 
\bal
\label{eq:SchEq}
-{T} \partial _t Y = -\frac{{T}^2}{2\Wi} \nabla ^2 Y + V Y\,, 
\eal
where
\bal
V &=\frac{{T}}{2}\nabla \!\cdot\! \left(\!\frac{\bm F}{\Wi} + \bm v\!\right)\! +\! \frac{\Wi}{2}\!\left(\frac{\bm F}{\Wi} + \bm v\!\right)^{\!\!2}\!\! + 2\eta \bm v \!\cdot\! \bm F.\!\!\!\!
\eal 
Note the $\eta \to -1 -\eta$ invariance of the potential. This invariance implies that the Gavallotti-Cohen fluctuation theorem holds \cite{LS,GC,GC1995aa}. 

The large time behavior is determined by the ground state energy $\lambda(\eta)$. We obtain the ground state energy for several different restitution forces in the following sections. 

\section{Results}
\subsection{Linear - Hookean elasticity}
The linear case, $\bm F = -\gamma \bm r$,  is integrable and corresponds to single particle quantum mechanics in a magnetic field \cite{landauQM}, where the ground state energy is 
\bal
\nonumber
\label{eq:Groundstate-Rouse}
\lambda(\eta) =& \frac{1}{\Wi} - \frac{1}{2\Wi}\left(\sqrt{(1+\Wi)^2 + 4 \Wi\eta}\right.
\\
&
\left.+ \sqrt{(1 -\Wi)^2 - 4 \Wi \eta}\right)
\eal
This expression holds for $\eta\in\left[-\frac{(1+\Wi)^2}{4 \Wi}, \frac{(1-\Wi)^2}{4\Wi}\right]$. Similar objects were derived in \cite{TCCP} and \cite{SMS}, where a polymer was placed in a shear flow. The Legendre transform of \EQ{Groundstate-Rouse} gives the LDF  
\bal
\label{eq:LDF-Rouse}
&\mathcal{L}(w) = [\eta _{-} w - \lambda(\eta_{-})]\theta(-w) + [\eta _{+} w - \lambda(\eta_{+})]\theta(w), \!\!\!
\eal
with
\bal
\nonumber
\eta_{\pm} =& -\frac{1}{2}\pm \frac{1}{4 \Wi w^2} \left(-3 +(1- (1 + \Wi^2)w^2)^2  \right.
\\
&\left.+ 2 \sqrt{1 + 2 (1 + \Wi^2)w^2}\right)^{1/2},
\eal
where $\theta(w)$ is the Heaviside step function. For large values of work the asymptotes are 
\bal
\lim_{w\to \pm\infty}\!\!\mathcal{L}(w) =\pm \frac{(1\mp \Wi)^2}{4\Wi}w\,. 
\eal
This implies that the PDF of the work is an exponential. Notice that for $\Wi > 1$ we have $\lambda(0) \neq 0$, which amounts to the breakdown of linear elasticity. Namely for strong velocity gradients the polymer can not be in a steady state if the restitution force is linear. This linear case analysis is straightforwardly generalizable to a $3d$ case. Below we focus on the nonlinear case. 

\subsection{Nonlinear elasticity}
For a general nonlinear force \EQ{SchEq} is non-integrable. However here $T$, the ratio between that temperature and the elastic energy at the maximal extension, is always smaller than unity, since we consider a nonlinear polymer in a steady state. Moreover often it is interesting to look at $T \ll 1$ which would mean that the natural length of the polymer spring is much smaller than its maximal length in the presence of the external flow. We refer to the regime $T \ll 1$ as the "semiclassical limit", due to the apparent analogy with quantum mechanics in \EQ{SchEq}. 
\begin{figure}[t]
\includegraphics[width=\columnwidth]{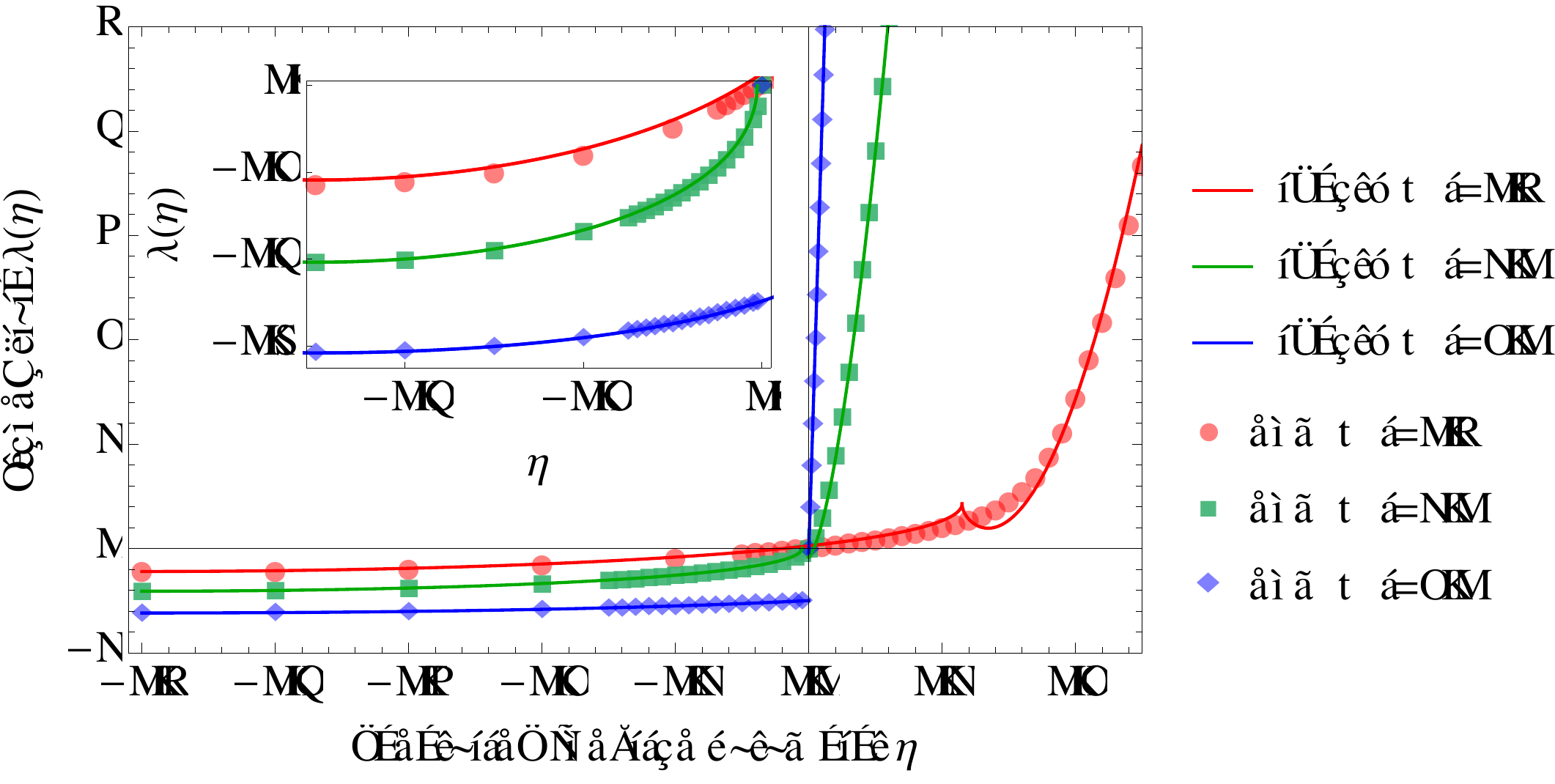}
\caption{\label{fig:rouse-fene-stitch-fig-a_v01.pdf} The ground state energy $\lambda$ of a FENE polymer as a function the generating function parameter $\eta$ at temperature $0.005\left(\gamma l^2/2\right)$. The inset zooms into the region $\eta \in [-0.5,0]$. The solid lines represent the semiclassical solution for the groundstate dominated by the root at origin  (see \EQ{Groundstate-FENEOrigin}) and the root in \EQ{root-unity}. The markers are the numerics done by a "Cloning algorithm" described in~\cite{GKP}.} 
\end{figure}
\begin{figure}[t]
\includegraphics[width=\columnwidth]{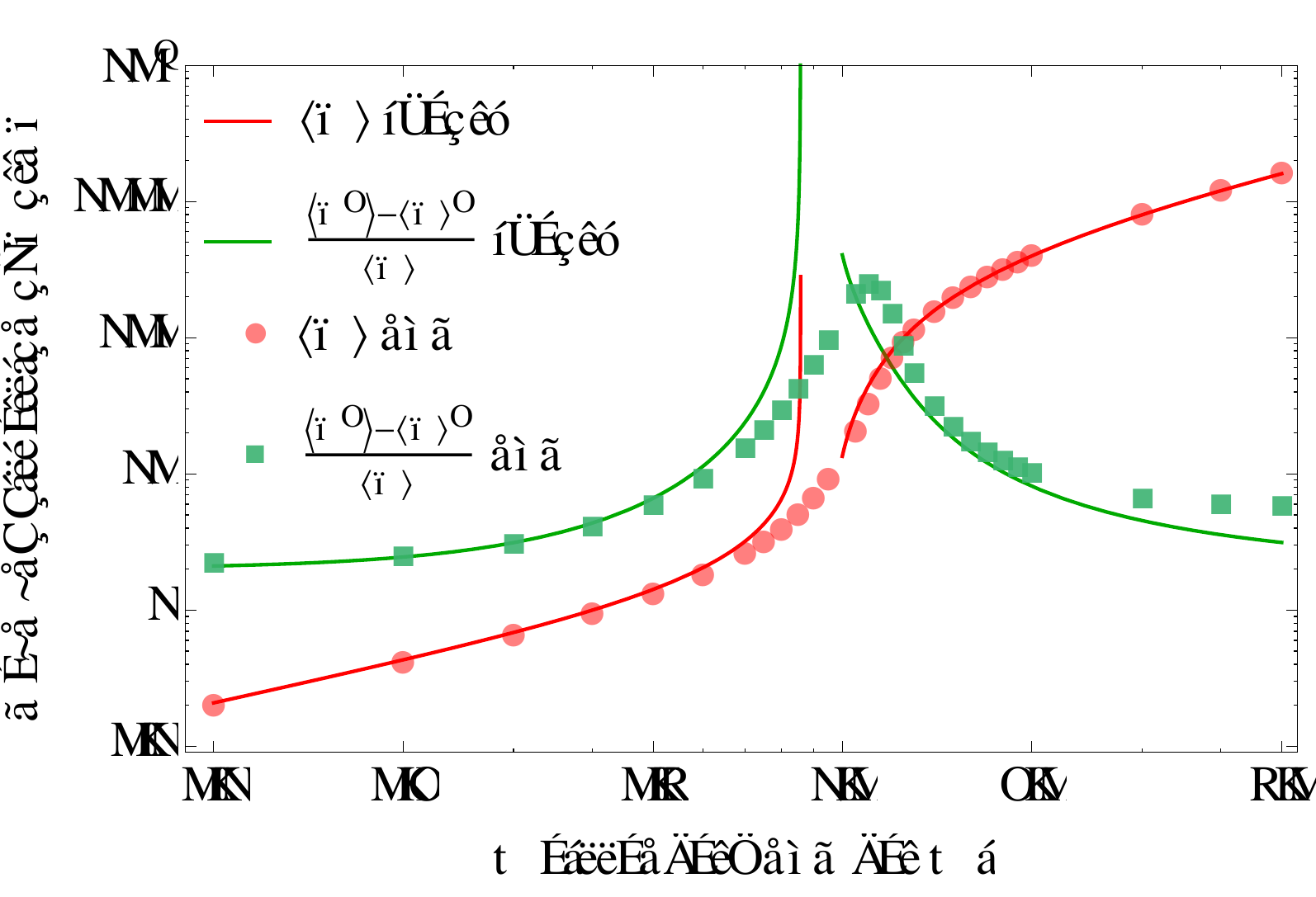}
\caption{\label{fig:mean_w_var_w_fene2d_v01.pdf}The mean and the dispersion of work $w$ for $\Wi = 0.1 \div 2.0$. The red solid line is $ \langle w \rangle= \lambda'(0)$, while the green solid line is $(\langle w^2 \rangle - \langle w\rangle^2)/\langle w \rangle = \lambda ''(0)/\lambda'(0)$. The ground state energy for $\Wi < 1$ is given by \EQ{Groundstate-FENEOrigin}, while for $\Wi > 1$ we have \EQ{Groundstate-semiclassic} at minimum \EQ{root-unity}. The dots represent the numerical estimates of the mean and dispersion obtained by averaging $10^5$ trajectories. In the simlations the evolution time was $10^3\tau$ and the temperature  was $0.005 \left(\gamma l^2/2\right)$.}
\end{figure}
Below we will describe an approximate way to obtain the ground state, $\lambda(\eta)$, for the FENE polymer. In the $T \ll 1$ regime we can assume that the polymer length is close to the minimum of the potential $V$. To find the ground state we expand the potential around the minimum $\bm r_*$ and add harmonic fluctuations 
\bal
\label{eq:Groundstate-semiclassic}
\lambda(\eta)\! \simeq -\frac{V(\bm r _*)}{{T}}\! -\! \frac{1}{2\sqrt{\Wi}}\!\left(\!\sqrt{V_{xx}(\bm r_*)} + \sqrt{V_{yy}(\bm r_*)}\!\right)\!.\!\!\!\!\!\!
\eal
The coupling term vanishes for $V$: $V_{xy}(\bm r_*) = 0$. Note that $\eta \to - \eta$ or $\bm v \to - \bm v$ changes the roles of $x,y$, also notice that this potential is symmetric around $x \to - x$ and $y \to -y$. Thus when searching for minima one can look at the e.g. $x > 0$ semi-axis to get the full picture. Depending on $\eta$, $\Wi$ and ${T}$ there are two deep minima, one at the origin, with ground state energy
\bal
\nonumber
\label{eq:Groundstate-FENEOrigin}
\lambda(\eta) =& \frac{1}{\Wi} - \frac{1}{2\Wi}\left(\sqrt{(1+\Wi)^2 + 4 \Wi\eta - 4 T}\right.
\\
&\left.+\sqrt{(1 -\Wi)^2 - 4 \Wi\eta - 4 T}\right)
\eal
valid for $\eta\in\left[-\frac{(1+\Wi)^2- 4 {T}}{4 \Wi}, \frac{(1-\Wi)^2- 4 {T}}{4\Wi}\right]$. The above expression differs from the linear case \EQ{Groundstate-Rouse} just slightly (terms with $T$). The other minimum is at $y_* = 0$ and 
\bal
&x_* = \left(1 + \frac{(2-4 {T})}{1 + 2\Wi + 4\Wi \eta} \sqrt{\frac{(1 +\Wi(2 + 4\eta))^3}{3\Wi^2(1 - 2 {T})^2}}\times\right.
\nonumber
\\
& \label{eq:root-unity}
\left.\cos\!\left(\!\frac{1}{3}\arctan\!\left(\!\sqrt{\frac{(1 +\Wi(2 + 4\eta))^3}{27 \Wi^2(1 - 2 {T})^2}-1}\right)\!- \!\frac{2\pi}{3} \!\right)\!\!\right)^{\!\!\!1/2}\!\!\!\!.\!\!\!
\eal
The ground state energy for $\eta\gg1$ can be approximated as $\lambda(\eta)\approx  (2 \Wi/ T) \eta^2$, and this leads to Gaussian statistics of $w$ (see \FIG{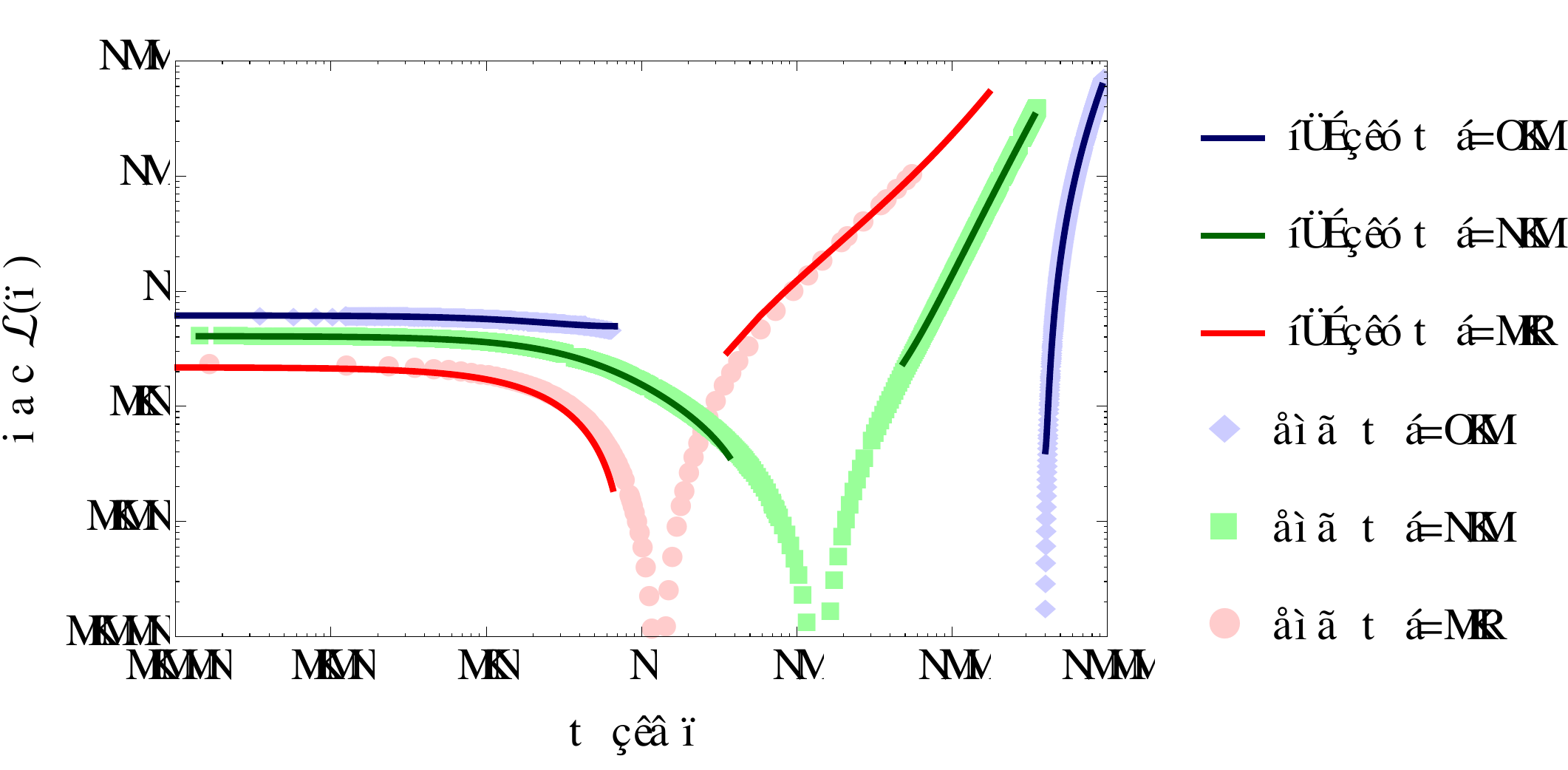}). The two different asymptotic are connected with an intermediate region, that we investigated numerically with a "Cloning algorithm"~\cite{GKP}. The results for the ground state energies are shown on \FIG{rouse-fene-stitch-fig-a_v01.pdf}. 

In an analogous manner one can consider different nonlinear forces, such as: $\bm F = - \gamma \bm r/(1-r)^2$ (worm like polymers \cite{MS}) and $\bm F = -\gamma \bm r /(1 - r^2)^{n}$. The formulas for the semiclassics at origin will be dominated by linear elasticity, e.g. in the later case we get \EQ{Groundstate-FENEOrigin} where we just need to replace ${T}$ with $n {T}$. In the nonlinear case large $\eta$ limit for $\bm F = - \gamma \bm r/(1-r)^2$ has a simple expression: $\lambda(\eta) \approx \left((1 + {T}) - T\sqrt{1-3 {T}}\right) (2 \Wi/T) \eta^2\,.$ The minimum of the corresponding potential is at $ \bm r_* \approx (1 - (2\Wi\eta)^{-1},0)$. Here the leading order with $T$ for $\lambda(\eta\gg1)$ is the same as for the FENE polymer. 

\subsection{Validity}
\begin{figure}
\includegraphics[width=\columnwidth]{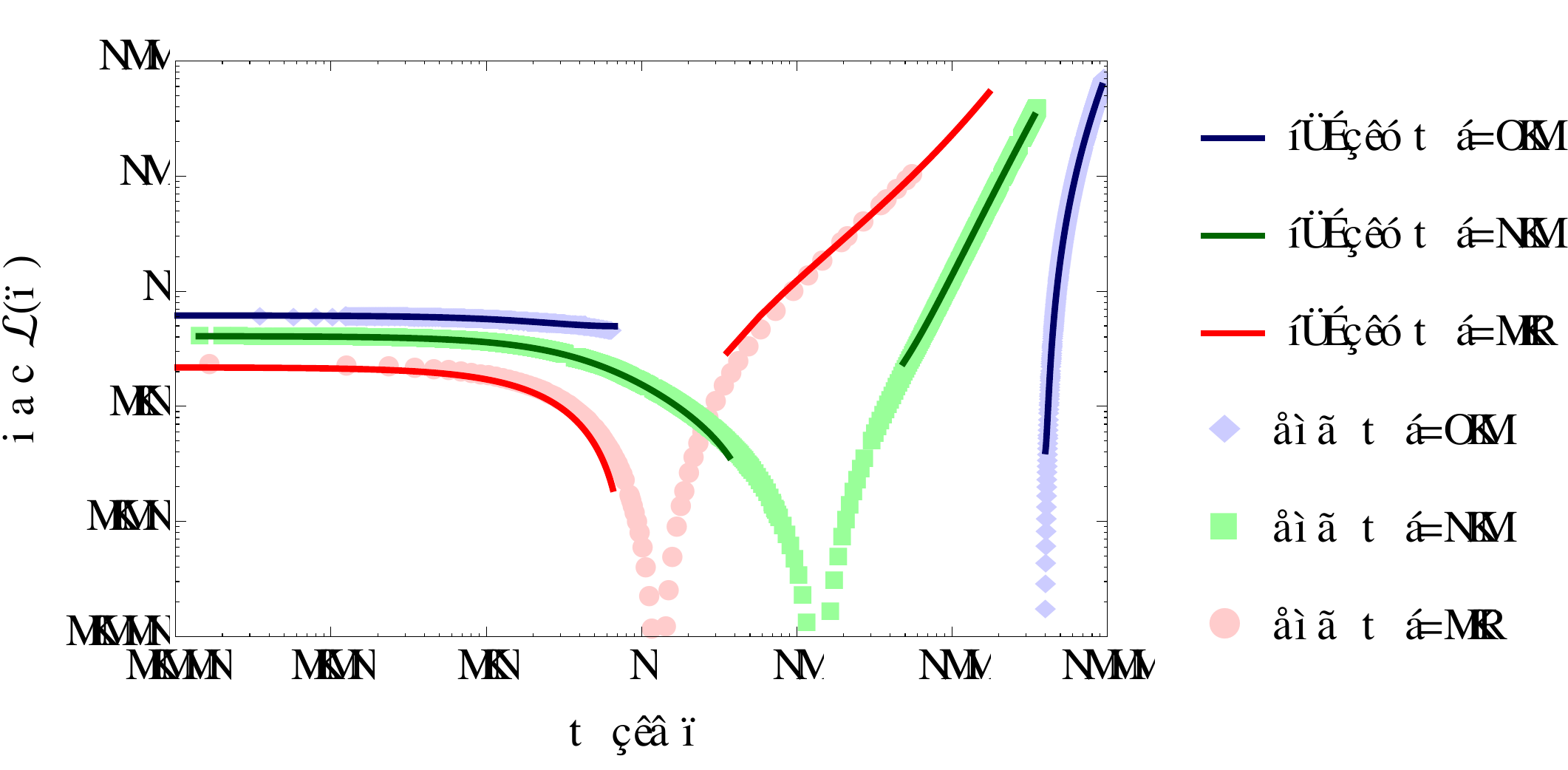}
\caption{\label{fig:LDF-FENE-both-v02.pdf}The LDF $\mathcal{L}$ as a function of work $w$. Here we show the two different assymptotics at temperature $0.005 \left(\gamma l^2/2\right)$. The markers represent the Legendre transform of the numerically obtained ground state energy $\lambda$ (numerics done with a "Cloning algorithm" described in \cite{GKP}). The solid lines represent $\mathcal{L}$ obtained analytically from the semiclassical solutions for the groundstate, given in \EQ{Groundstate-semiclassic}, dominated by root at origin (\EQ{Groundstate-FENEOrigin}) and the root at \EQ{root-unity}.}
\end{figure}

In the considered cases the ground state energies were continuous and convex. Therefore the Gartner-Ellis theorem is applicable and it guaranties that the LDF is the Legendre transform of the ground state energy \cite{Ellis}. The LDF of ground state energies (shown on Fig.~\ref{fig:rouse-fene-stitch-fig-a_v01.pdf}), can be seen on Fig.~\ref{fig:LDF-FENE-both-v02.pdf}.

The semiclassical description holds as long as the semiclassical ground state wave function $|Y_g\rangle$ width is smaller than the system size: 
\bal
\max\left[1/\sqrt{V_{xx}(\bm r_*)}, 1/\sqrt{V_{yy}(\bm r_*)}\right]\ll\sqrt{\Wi}/{T}
\eal
 and the kinetic term in \EQ{SchEq} is negligible compared to the potential part, i.e. 
\bal
\langle Y_g |-({T}^2/2 \Wi)\nabla^2 |Y_g\rangle\ll \langle Y_g | V |Y_g \rangle.
\eal For FENE polymer at $T = 0.005 \left(\gamma l^2/2\right)$  the semiclassical description is a good approximation almost everywhere: for $\Wi = 0.5$ it works for $0.1>\eta > 0.12$ and  $\Wi = 1.0\div 2.0$ it works everywhere except in the vicinity of $\eta = 0$ (c.f. to Fig.\ref{fig:rouse-fene-stitch-fig-a_v01.pdf}). Our simulations of the semiclassical ground state $\lambda(\eta)$ (see \FIG{rouse-fene-stitch-fig-a_v01.pdf}) were done by a "Cloning algorithm"\cite{GKP}. The parameters of the simulation were: time-step $0.01 \tau $ and evolution time $10^3 \tau$.

Especially it is interesting to look at the phase transition at $\Wi = 1$. Notice that the ground state energy is discontinuous at $\eta = 0$ for $\Wi > 1$. We use our analytical expressions for the ground state energy $\lambda(\eta)$ to find the mean $\langle w\rangle = \lambda'(0)$ and the dispersion of work $(\langle w^2\rangle - \langle w \rangle^2)/\langle w \rangle = \lambda '' (0)$, in the vicinity of $\Wi = 1$. The analytical results away for the transition match the Monte Carlo averages over the polymer trajectories \FIG{mean_w_var_w_fene2d_v01.pdf}. Notice that the dispersion goes to a maximum at $\Wi = 1$. This corresponds to the multitude of very different polymer configurations that are present at the transition. Below the transition, $\Wi \ll 1$,  $\langle w \rangle \propto 2 \Wi$, $(\langle w^2\rangle - \langle w \rangle^2)/\langle w \rangle \propto 1/\Wi$. Close to the transition $\Wi \to 1^{-}$ we have $\langle w \rangle \propto 1/(1 - \Wi)$ and $(\langle w^2\rangle - \langle w \rangle^2)/\langle w \rangle \propto 1/(1-\Wi)$.

\section{Discussion and Conclusions}
It is important to emphasize that our theory and numerics work well for flows of different gradients strengths, as our assumptions only require small $T$ (small thermal fluctuations), and $T$ is a flow independent parameter. In the "semiclassical" limit (small $T$) the nonlinear dumbbell spends most of the time in the "coiled" or in the "extended" configurations. The drag coefficient for long time intervals is or that of a sphere or that of a thin rod, respectively. Thus, albeit simple and ignoring hydrodynamical interactions, our model provides important insights into the statistics of work and dissipation of polymers in gradient flows. 

We wish to highlight that even for nonlinear systems it is often possible to theoretically investigate objects like the LDF. Rare events corresponding to anomalous rate of entropy or work production are related to particular configurations of the polymer molecule. It can be especially insightful to look at the LDF near phase transitions, where its landscape is richer, due to the occurrence of different phases and many configurations that the system can take. In particular, experimental results on the statistics of work of stretching of polymers, near the coil-stretch transition, show critical slowing down and enhanced fluctuations \cite{08GS}. These effects, as the authors of the experimental study \cite{08GS} argue, most likely occur due to the presence of a large number of possible polymer configurations in the vicinity of a continuous thermodynamic phase transition. In addition, one could use LDF statistics to discern between different restitution forces. For the commonly used singular potentials describing the finitely extensible polymers, our results show that the LDF does depend on the shape of the potential.

Modern experimental techniques allow one to track single polymers. Dynamics of polymer molecules in external flows was extensively studied, see e.g.~\cite{PSC, PSCbook}. Such experiments improved the understanding of mechanical properties of polymer molecules. Measurement of the work production provide another way of approaching the same problem, such measurements could test our LDF results (see \cite{2013Latinwo, Latinwo2014aa}). Also by variation of the external flow one could study the polymers in coiled and stretched states.

The situation considered in this letter is quite general. We believe that our methods and results can be used in as a probe of soft matter dynamics in other systems, such as various nano-devices, molecular motors, polymer solutions, etc. Possible experimental realizations include elastic turbulence, drag reduction, optical tweezers experiments, etc. 

\section{Acknowledgements}
The authors acknowledge illuminating discussions with T.~Witten, A.~Grosberg, S.~R.~Varadhan, L.~Zdeborova, F.~Krzakala, and L.~Peliti, and fruitful comments made by the referees. The work at LANL was carried out under the auspices of the National Nuclear Security Administration of the U.S. Department of Energy at Los Alamos National Laboratory under Contract No. DE-AC52-06NA25396. MV thanks the Aspen Center for Physics and  the NSF Grant \#1066293 for hospitality during the preparation of this manuscript.

\bibliography{fdt-polymer-v15-draft.bib}

\end{document}